# Datamining a medieval medical text reveals patterns in ingredient choice that reflect biological activity against the causative agents of specified infections

**Erin Connelly***

Schoenberg Institute for Manuscript Studies, University of Pennsylvania, Philadelphia, PA 19010

erincon@upenn.edu

**Charo I. del Genio**

School of Life Sciences, University of Warwick, Coventry, CV4 7AL, UK

c.i.del-genio@warwick.ac.uk

**Freya Harrison**

School of Life Sciences, University of Warwick, Coventry, CV4 7AL, UK

f.harrison@warwick.ac.uk

## Abstract (150 words)

The pharmacopeia used by physicians and lay people in medieval Europe has largely been dismissed as placebo or superstition. While we now recognise that some of the *materia medica* used by medieval physicians could have had useful biological properties, research in this area is limited by the labour-intensive process of searching and interpreting historical medical texts. Here, we demonstrate the potential power of turning medieval medical texts into contextualised electronic databases amenable to exploration by algorithm. We use established methodologies from network science to reveal statistically significant patterns in ingredient selection and usage in a key text, the fifteenth-century *Lylye of Medicynes*, focusing on remedies to treat symptoms of microbial infection. We discuss the potential that these patterns reflect rational medical decisions. In providing a worked example of data-driven textual analysis, we demonstrate the potential of this approach to encourage interdisciplinary collaboration and to shine a new light on the ethnopharmacology of historical medical texts.

Infection remedies from medieval Europe often use ingredients known to possess some antimicrobial or immunomodulatory effects, at least in *in vitro* assays [1, 2]. Further, these texts suggest complex preparations of multiple ingredients, and contingencies of treatments for the same symptoms. A recent study found that an Anglo-Saxon remedy for eye infection used ingredients that when combined produced a strong antibacterial cocktail [3]. The existence in the medieval pharmacopeia of groups of ingredients that are repeatedly combined to treat the symptoms of infection, and that, individually, have useful biological activity, gives evidence to the hypothesis that medieval medicine had some rational basis [4, 5]. It also suggests that these remedies may contain combinations of natural materials that could yield clinically-useful biologically active molecules.

Network science is a powerful tool to model complex systems and extract meaningful information from data sets representing interactions between elements, such as the co-occurrence of ingredients in recipes. Network analysis has been used to characterise and analyse interactions between people or animals [6, 7, 8, 9], computer networks [10], gene regulation networks [11] and combinations of ingredients in culinary recipes [12]. The network approach represents a system as a graph, in which the nodes correspond to discrete elements (individual organisms, computers, genes or ingredients), and the links to dyadic relations between them (communication, coordinated behaviour or co-localisation) [13, 14, 15, 16]. The network paradigm provides researchers with mathematical tools to define and study quantitative properties of real-world systems, such as the typical distribution of links amongst nodes [17, 18], the relation between the number of links of the nodes sharing a link [19, 20], the ability of nodes to synchronize their dynamics [21, 22], and the existence of communities [11, 23, 24]. In this context, communities (also known as modules) are groups of nodes whose density of internal connections is higher than that of the links between different modules [25].

Community structures have been identified in numerous real-world systems, where the module elements often share a common function. Some typical examples are closer friendship groups within sport clubs [26], ensembles of closely collaborating jazz players [27], and sets of chemical substrates that participate in the same metabolic reactions [28, 29]. Detecting such modules is one of the main techniques that are routinely employed to create organic descriptions of complex systems; a large number of algorithms have been created to perform this task [23, 24, 30, 31, 32]. Here, we use community detection to analyse a set of medieval medical remedies and discover communities of ingredients that are commonly combined to treat symptoms of microbial infection.

Our work has two goals. First, we aimed to determine the tractability of turning medical texts into contextualised electronic databases amenable to quantitative analyses of hidden patterns in the use and preparation of historical *materia medica*. Network analyses appear eminently suited to interrogating medical texts, and have successfully been used to analyse patterns of ingredient co-occurrence in culinary recipes [12]. We therefore wished to determine whether community detection procedures could be used with a medieval medical text, given the difficulties of working with medieval manuscripts (orthography, linguistic change and the need for contextualisation of language based on a knowledge of the medical and material culture of the period). Our second goal was to apply community detection algorithms to determine whether an exemplar text contained detectable patterns of ingredient combination and/or assignation of specific ingredients to specific symptoms of disease (in our case, microbial infection), and consider, based on the ethnopharmacological literature, whether these patterns could reflect rational medical decision-making on the part of the medieval physician.

The limited availability of medieval medical texts, as well as the constraints of researchers mining these texts by hand, has thus far meant that ethnopharmacological research conducted on medieval European medicine has concentrated on single specific ingredients (e.g. a body of literature on *Plantago* spp. [33, 34, 35]), or single recipes from select medieval texts [3, 36]. The use of digital technologies to turn these texts into databases amenable to quantitative datamining requires a careful interdisciplinary approach, but it could provide an entirely new perspective on medieval science and rationality. It could even suggest combinations of *materia medica* worthy of further study to determine their potential for drug development.

We chose as our exemplar text the fifteenth-century Middle English manuscript known as the *Lylye of Medicynes*. This is a Middle English translation of Bernard of Gordon's *Lilium medicinae*, originally completed in 1305. The text is extant in one manuscript, Oxford, Bodleian Library MS Ashmole 1505. Bernard of Gordon was a significant medieval medical doctor and lecturer in Montpellier [37, 38]. His magnum opus is a lengthy treatise on disease etiology, medical philosophy and history (citing Galen, Hippocrates, Ibn Sīnā, and many others), personal case studies, and treatment procedures and recipes. The first owner of the *Lylye of Medicynes* is thought to have been Robert Broke, apothecary to Henry VI; appropriately, the Middle English translation is notable for its pharmaceutical content [39, 40, 41]. There are 360 recipes in the *Lylye* specifically notated with *Rx*, as well as nearly 6,000 ingredient names throughout the whole text. The recipes in the *Lylye* are presented in a standard format: they begin with an indication of what type of remedy it is, such as ointment, syrup, plaster, and usually the phase of the illness during which it should be applied (beginning, middle, or end). This is followed by a recipe of ingredients, often accompanied by quantity, and preparation instructions, such as to boil, powder, or infuse. In the *Lylye* and other medieval medical texts, it is common practice to record recipes without defined measurements, and to include long lists of ingredient substitutions instead of exact quantities, even when hazardous materials are involved. Often, specific details are not given because it is assumed that the reader is already familiar with the methodology. Furthermore, some diseases or treatments carried so many variables that the selection of ingredients and quantities was left to the judgement of the physician in charge of the patient.

The *Lylye* is a unique translation of a significant text by a notable medieval physician. Its pharmaceutical content, association with a medieval apothecary and Tudor barber-surgeons, large number of ingredients, and chapters on infectious disease are all factors which make this text an attractive starting off point for an analysis of significant antimicrobial ingredient combinations.

## Results

*Turning the* Lylye of Medicynes *into an electronic database*

As a starting point, the 360 recipes of the *Lylye* were entered into a spreadsheet (**Supplementary Data**). These recipes contain 3,374 individual ingredient names for the treatment of 113 medical conditions. Of those disease conditions, 30 (containing 86 recipes) may be classified as external infections, mainly of the skin, mouth, or eyes. The *Lylye of Medicynes* uses language which clearly indicates *infection*, as it might be translated in modern terms. For example, in describing skin swellings, abscesses, lesions, wounds, carbuncles, erysipelas and fistula, symptoms from the text include broken skin, purulence, itching, foul smell, heat, moisture, aching, pricking and burning sensations, redness, yellowness, ulceration, and black crusts, which all are indicators of infection (Book 1, fols 24r-28v).

In addition to the 360 recipes identified by the text with Rx, there are multiple other ingredients, which are not prefaced by Rx, but clearly are set up in the format of a recipe. To the principal data set, 36 additional recipes for the treatment of 19 medical conditions (specifically for external infection) containing 174 ingredients were added for a total data set of 3,548 ingredients (747 unique names) and 124 unique disease names, of which 41 may be classified as potential external infections. Ingredients which were excluded from the data set were those not in recipe format, such as lists of ingredient substitutions and lists of purgative ingredients.

There are numerous challenges to constructing and processing a medieval data set using modern analytical tools. These challenges include medieval spelling and language variation, multiple synonyms for the same ingredient, translation of medieval ingredients and diseases into modern equivalents (many of these terms have multiple possible interpretations or ambiguous definitions), and the variation within the modern system of botanical binomial nomenclature. Middle English spellings may be standardized and hand-checked using the Vard 2 software [42, 43, 44]; however, this program was developed specifically for Early Modern English spelling variation and was deemed not appropriate for our data set from the Middle English *Lylye of Medicynes*. Spelling variants were normalized by hand with preference given to the English word (e.g. *oil* instead of *oleum*). Additionally, the frequency of occurrence in the text of the *Lylye* and the headword used in the *Middle English Dictionary* (*MED,* [45]) were taken into consideration when normalizing the spelling. The Middle English characters of thorn, yogh, and long *z* were modernized. The Middle English usage of *u/v* and *i/j* were also standarised to modern practices.

R. James Long [46] has stated that in translations of Latin botanical terms 'as many as 15 synonyms for the same plant have been counted and it is not unheard of that two or more words for the same herb appear in the same recipe.' This is an accurate description of ingredient terminology in the *Lylye*'s recipes. To eliminate redundancy and to aid the processing of our data set, synonyms were combined under one term where possible with preference given to the Middle English term appearing most frequently in the *Lylye* or as the headword in the *MED*. For instance, *fenel* was chosen as the headword for textual appearances of *fenel*, *fenell*, *feniculi*, *feniculum*, *marathri*, *maratri*, *maratrum*. Due to variations in active ingredients in different parts of a plant, those designations were included (e.g. root of fenel, seeds of fenel, juice of fenel). The imperative of this study was to identify significant ingredient combinations. Therefore, preparation instructions were not included with the ingredient names, except in cases where the preparation changed the nature of the base ingredient [e.g. *calx* (lime); *calx vive* (calcium oxide); *calx extinct* (calcium hydroxide)]. The Middle English ingredient names and disease names were not translated into Modern English in the spreadsheet or for the analyses. Modern English translations are provided for the ingredients referenced in this article. As many medieval medical terms are ambiguous and definitions vary between modern dictionaries, caution was used in translating these terms into Modern English and in identifying the scientific names of plants. The definitions were cross-checked using the *MED*, *Oxford English Dictionary* [47], the *Dictionary of Medical Vocabulary in English, 1375-1550* [48], Kew Medicinal Plants database [49], and the Plant List database [50].

*Network analysis of ingredients in the* Lylye of Medicynes

To build a network from the recipes, we created a node for each individual ingredient encountered. Every time two ingredients were found together in the same recipe, a link was placed between them. In cases where pairs of ingredients were found in more than one recipe, the corresponding links were "strengthened" by assigning them a weight proportional to the number of times the pair occurred across recipes. As an example, Fig. 1 shows the network created from one recipe for *fistula in lacrimali* (lacrimal fistula) and one for *pascionibus oris* (diseases of the mouth). The ingredients for the former (shown in yellow) are *galle* (bile salts), *hony* (honey), *pomegarnettes* (pomegranates), *ruta* (a plant of the genus *Ruta*, especially common rue, *Ruta graveolens*), and sumac (sumac, *Rhus coriaria*); those for the latter (shown in blue) are *galle*, *hony*, *olibanum* (frankincense), *pomegarnettes*, *sumac*, and *vinegre* (vinegar). Note that four ingredients, namely *hony*, *galle*, *pomegarnettes*, and *sumac*, are found in both recipes. Thus, the corresponding nodes are colored both yellow and blue. Since, in this example, six pairs of ingredients are found in both recipes, the links that represent them are thicker. When this procedure was carried out on all the recipes, we obtained a network with 354 nodes and 3073 weighted links.

> **Figure 1** Example of an ingredient network. The nodes in yellow are ingredients of a recipe for *fistula in lacrimali*; those in blue are for *pascionibus oris*. Ingredients found in both recipes are colored both yellow and blue. Thicker links join pairs of ingredients that appear in both recipes.

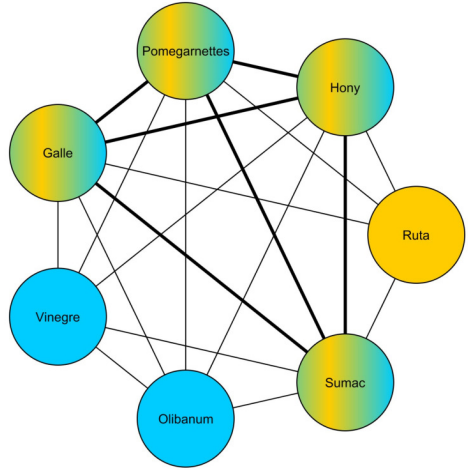

To normalize the link weights, we used a linear transformation that did not affect the network properties [51, 52]. More specifically, if nodes *i* and *j* were connected, their link weight $W_{i,j}$ transformed as

$$W_{i,j} \to \frac{(1-2\varepsilon)W_{i,j}+(2\varepsilon-1)W_{min}}{W_{max}-W_{min}} + \varepsilon \qquad (1)$$

where $W_{min}$ and $W_{max}$ are the minimum and maximum original weights, and $\varepsilon$ is a small parameter used to avoid cutting links whose original weight is $W_{min}$. In our case, we used the common choice $\varepsilon$ = 0.01. In principle, one could perform the analyses of ingredient communities on the full weighted network. However, the results could be affected by noise in the data, such as could be originated by pairs of ingredients whose links were too weak to be significant, but strong enough to spuriously alter the local structure of the network. To avoid this type of problem, we introduced a variable threshold *t*. Starting from the full weighted network, for every given value of *t* we created a thresholded, unweighted network by removing all the links whose weight was ≤ *t* and considering all other links as having equal unitary weight. By studying how the communities change with the value of the threshold, we were able to find groups of ingredients that always or most often belonged to the same community, regardless of the assignment of the other nodes around them. This allowed us to identify combinations of ingredients that were statistically relevant and thus warranted further investigation.

To illustrate the procedure, consider the fictitious network in Fig. 2. For threshold 0, we identify two communities (red and blue), while for thresholds 0.1 and 0.2 we find three communities (red, blue, and violet). Note that, in going from threshold 0 to 0.1, the blue community splits into two sub-communities (blue and violet). This is an example of hierarchical structure: some communities are themselves composed of other, smaller communities. Comparing the results for the three thresholds, we find the combinations of nodes 1 and 3, nodes 5, 6 and 7, and nodes 10 and 12 to be important. Note that, in the original network, nodes 5 and 6 are connected with the weakest (thinnest) link. Nonetheless, this procedure identifies their co-occurrence, together with ingredient 7, as potentially relevant.

**Figure 2** Fictitious example of the process for identifying relevant combinations of ingredients via thresholding and community detection. (A) The starting network. (B, C) Apply thresholding procedure and choose partitioned network with maximal modularity ($q_{\{c\}}$, equation 2). (D) Draw heatmap to visualise strengths of associations between ingredient pairs. Darker pixels indicate stronger association. For example, in this example network ingredients 1 and 3 are more strongly connected by their recipe co-occurrences than are ingredients 1 and 2 or 2 and 3. (E) Identify most strongly associated group of ingredients and manually search database for recipes including these combinations.

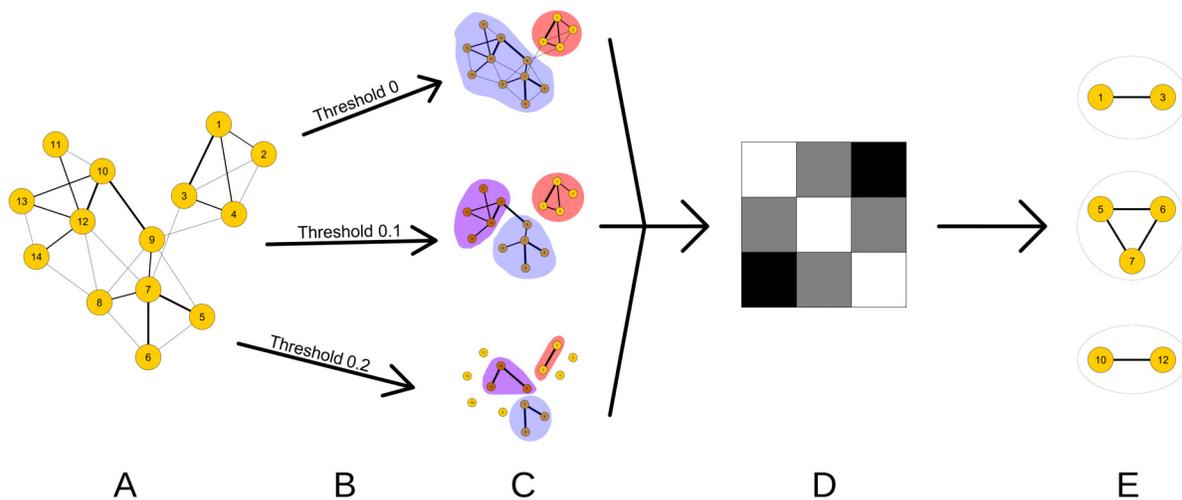

We chose 6 thresholds: 0, 0.01, 0.15, 0.29, 0.43 and 0.571. Our choice was based on the fact that the number of edges that are cut increases with $t$, causing more nodes to become isolated, and the network to fragment (see Fig. 3). Thus, we chose the values at which such changes in the thresholded network happened, stopping at 0.571 because for that threshold the number of non-isolated nodes becomes only 10.

To carry out the community detection on our data, we used the algorithm developed by Treviño et al.[32]. This is a spectral method that uses a synergy of refining steps, including local and global tuning, as well as agglomeration, to estimate the network partition that maximizes modularity, a widely used objective function. For a network with $N$ nodes and $m$ links, its partition into a set of communities has modularity

$$q_{\{c\}} = \frac{1}{2m}\sum(A_{i,j} - \frac{k_i k_j}{2m})\delta_{c_i c_j} \qquad (2)$$

In the expression above, $A$ is the adjacency matrix of the network, whose elements $A_{i,j}$ are 1 if nodes $i$ and $j$ are linked, and 0 otherwise; $k_i$ is the number of links involving node $i$, $c_i$ is the community to which node $i$ belongs, $\delta$ is Kronecker's symbol; and the sum is over all possible pairs of nodes. Thus, this definition promotes partitions in which the number of links within communities is higher than its expectation value for a randomized version of the network.

> **Figure 3** Fragmentation of the network with threshold. As the threshold is increased, more nodes lose all their links and become isolated.

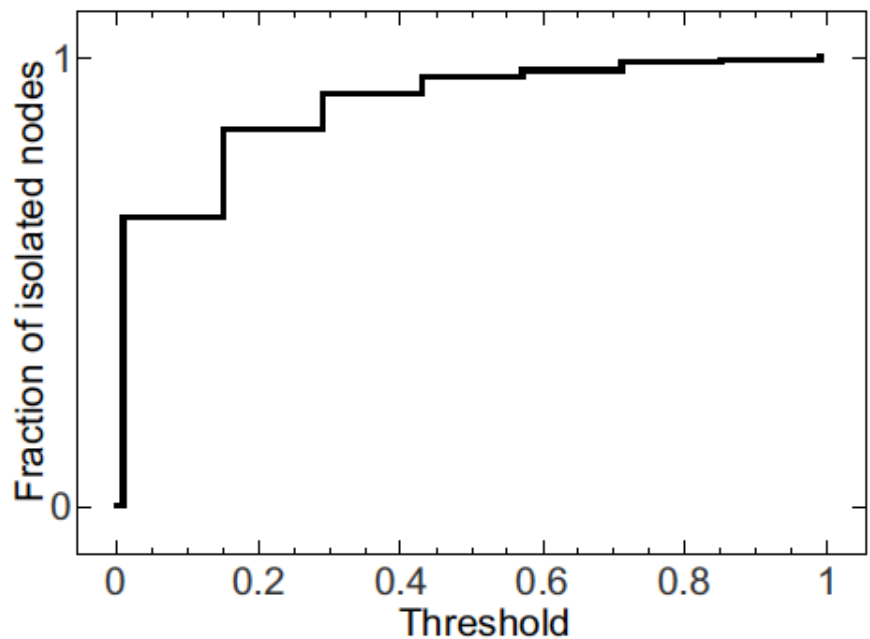

*Communities of ingredients in the* Lylye of Medicynes

For each value of link threshold, we ran our community detection algorithm 10,000 times, and selected the partition with the highest modularity. Having obtained the best community structures, we counted the number of times every possible pair of ingredients was assigned to the same community, producing a matrix of co-occurrences. To visualize this matrix, we sorted the ingredient list using the Cuthill-McKee algorithm [53], and represented the result as a greyscale heat map, in which the saturation of each pixel of coordinates ($i, j$) is proportional to the number of community co-occurrences of ingredients $i$ and $j$ (Fig. 4).

The results clearly show the existence of a hierarchical structure within the recipes. Each larger community is composed of other, smaller sub-communities, all with a common kernel of ingredients. This allows us to identify three core combinations and four core individual ingredients. The combinations are *aloen* (aloes; *Aloe vera*)+*sarcocolla nutrite* (gum which exudes from one of several Persian trees, including *Penaea mucronata* and *P. sarcocolla*, *Astragalus* *fasiculifolius* and *A. sarcocolla*; *nutrite* means infused, typically in breast milk); *ceruse lote* (white lead, washed or purified)+*eris ust* (copper calcined), and *olibanum*+*sumac*. The single ingredients are *balaustia* (blossoms of pomegranate), *galle*, *hony*, and *vinegre*. Of these, we excluded *ceruse* and *eris* from our consideration. Although certain metals, such as copper, have shown antimicrobial properties, medicinal plants and plant-derived products were the main focus of this enquiry. Thus, we focus on the remaining ingredients.

**Figure 4** Community co-occurrence of ingredients. Every pixel in the figure corresponds to a pair of ingredients. The saturation of the color is proportional to the number of times the pair of ingredients is found in the same community.

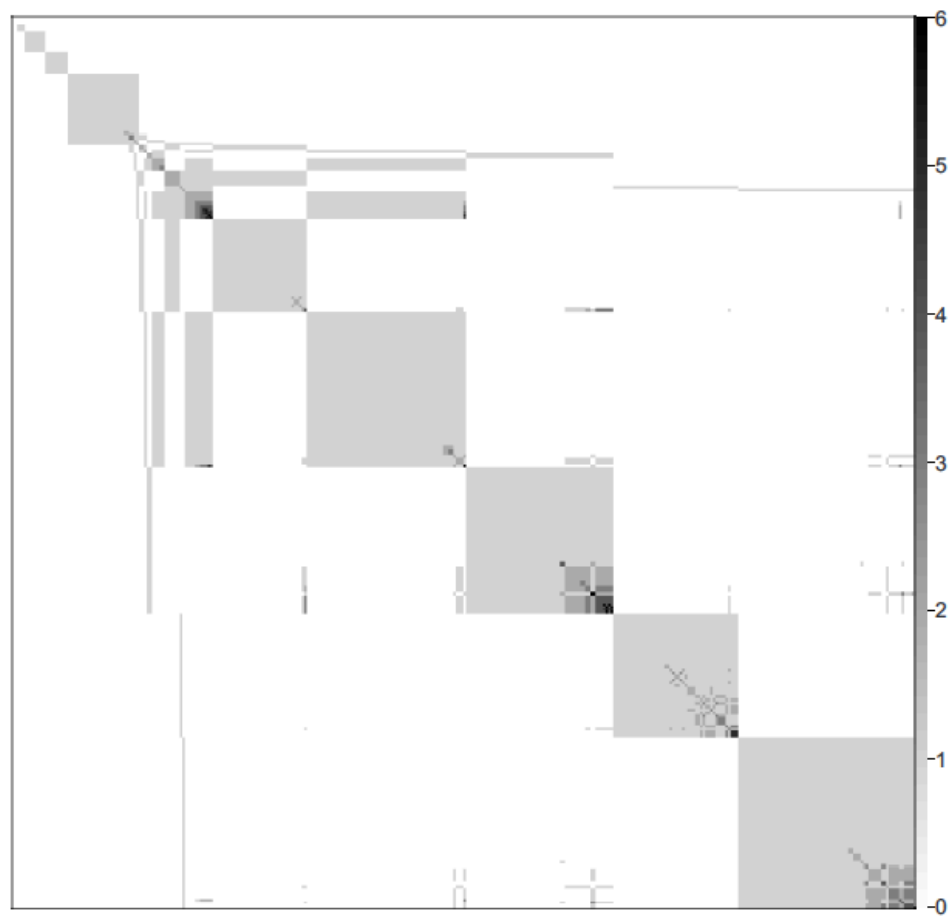

*Choice of two exemplar remedies*

From the data set, 62 recipes were manually selected as candidates for further investigation based on the results of network analyses. These 62 recipes were placed into seven groups in order of potential relevance (number of ingredients or pairs of ingredients that occur in the recipe; **Supplementary Data**). Remedies containing several alchemical metals, opiates, or lengthy, complex lists of ingredients (e.g. one recipe contains 32 ingredients) were excluded. The two most interesting remedies were selected based on the exclusion criteria, the antimicrobial significance of the ingredients as individuals and as combinations, and the description contained in the text of the condition they were intended to treat. These two recipes were compared with an early printed edition of the Latin *Lilium medicinae* (Lyon 1559) to see if there was any significant variation in the recipes between texts.

The first recipe occurs in a chapter for the treatment of eyes, including many infectious conditions. This particular recipe is for the treatment of *fistula in lacrimali*, which may be a lacrimal fistula, by modern definition; it is described by the text as 'a fester in the corner of the eye' (fol. 99r). This recipe is described as an ointment to 'regendre flesch' (i.e., to cause new flesh to grow/to repair/to heal). This recipe states to make an ointment of *aloen*, *olibanum*, *sarcocolla [nutrite]*, *sumac*, and *rind of pomegarnettes*. The inclusion of *nutrite* was informed by the Latin text and by the frequent occurrence of *sarcocolla nutrite* throughout the Middle English *Lylye*. It is possible that it was omitted in this recipe by scribal error or an errant (or variant) exemplar.

The second recipe occurs in the chapter *De pascionibus oris* for the treatment of 'pustules, ulcers, *apostemata* (swelling/inflammation), cancer, fistula, *herpestiomenus* (gangrene), and *carbunculus* (carbuncle; suppurating boil)' (fol. 126r). This recipe is a *gargarisme* (a medicine for washing the affected area of the mouth or throat) to be used if the condition is deemed to be caused by the phlegmatic humour. The mouthwash is to be made of *sumac*, *galle*, *psidia* (the rind of pomegranate or the bark of the tree), *balaustia*, *mastic* (resin exuded from the mastic tree,

*Pistacia lentiscus*), *olibanum*, *hony*, and *vinegre*. We suspect it likely that these two recipes are especially good candidates for further research.

*Structured review of evidence for efficacy of selected ingredients.*

To perform a thorough literature search for evidence of antimicrobial or immunomodulatory effects of the selected ingredients, we (a) searched the Cochrane Database of Systematic Reviews (www.cochranereviews.com) for reviews containing each ingredient's name, and (b) conducted a wider literature search using Scopus.com (*TITLE-ABS-KEY (ingredient_name) AND TITLE-ABS-KEY (antibacterial OR antifungal OR antibiotic OR antimicrobial)*). For plant species, common names and Latin binomials were searched; further terms were added as necessary to specifically search for active compounds within ingredients (e.g. lactoferrin for breast milk). Searches were conducted during the week of 25-29th October, 2017. In empirical literature on the antimicrobial effects of ingredients of component compounds, we assessed the rigour and relevance of the results. This included not only a judgement on whether sufficient quantitative data were presented, but also an assessment of the relevance of experiments to an *in vivo* context, i.e., we judged experiments testing the effects of substances on bacteria growing as biofilm [54] as more relevant than those using disk diffusion assays on agar plates or inhibition of planktonic growth, and those using clinical isolates of bacteria as more relevant than those focusing purely on domesticated lab strains of the same bacteria.

We found poor evidence for useful antimicrobial or healing effects of aloe, frankincense, mastic, and sarcocolla. Aloe extracts are used in a wide variety of medicinal and cosmetic products marketed for skin healing, and numerous patents cover the use of acetylated mannans from *Aloe vera* due to their supposed immunomodulatory effects [55, 56]. However, two recent Cochrane reviews found that published studies on the wound healing and antiseptic properties of this plant were generally of low quality and that the results published of human trials returned variable results, spanning help and harm to the patients [57, 58]. The available literature on the biological composition and activity of mastic was even less clear. This resin has a history of use as a chewing gum to promote oral health, and does appear to contain compounds with the ability to inhibit bacterial growth in disk diffusion assays *in vitro* [59]. However, we found that the few publications available which specifically addressed the effect of mastic gum on oral bacteria did not present clear quantitative data on the effects of mastic on bacterial cell viability [60, 61, 62, 63]. Frankincense may be obtained from a number of trees in the genus *Boswellia*, most usually *B. sacra*. It may have some immunomodulatory and thus healing effects via its effects on the NF-κB and TNFα pathways, but little research has been conducted on its chemistry; this is further complicated by the fact that different species have different chemical compositions and crude extracts can have different effects on inflammatory pathways than purified constituent molecules [64, 65]. Finally, while medicinal use of sarcocolla is mentioned in Latin, Greek, and Arabic historical sources, we were unable to find any modern research into its biological properties.

Slightly better evidence was found suggesting antibacterial potential of vinegar, sumac, bile, and pomegranate rind. Vinegar is generally acknowledged to act as a mild disinfectant, and pure acetic acid has been shown to exhibit various bactericidal effects at similar or lower concentrations to those typically found in vinegar. This includes the ability to kill cells of key opportunistic pathogens living as monospecies biofilms (clinical isolates of the Gram-negative species *Escherichia coli, Pseudomonas aeruginosa, Acinetobacter baumanii, Klebsiella pneumoniae, Enterobacter cloacae* and *Proteus mirabilis*, plus the Gram-positive *Staphylococcus aureus* [66, 67]. As a weak acid, it can readily cross cell membranes and collapse the cross-membrane proton gradient necessary for ATP synthesis; once inside the cell, acetic acid alters the cytoplasmic pH and this can cause DNA damage and protein unfolding [67]. A clinical trial into the effectiveness of acetic acid for treating burn wound infections is underway in the UK [68]. Vinegar has also been shown to kill the fungus *Candida albicans*, the causal agent of oral thrush, *in vitro* [69]. Like vinegar, an aqueous extract of sumac is acidic [70]; this property in itself could provide some antibacterial effect, but even after increasing the pH to neutrality, aqueous extracts of dried sumac fruit show bactericidal effects against Gram-positive and Gram-negative bacteria grown as planktonic cultures, and this has been partly attributed to tannins [70, 71]. The bactericidal effects of bile are long established: bile salts disrupt bacterial membranes and this quality contributes to host-mediated control of bacterial overgrowth in the gut [72, 73]. Interestingly, different fractions of aqueous or alcohol extracts of pomegranate peel have different spectra of antibacterial activity as measured in disk diffusion assays, and different authors present conflicting data on the inhibitory effects of aqueous extract fractions [74, 75, 76, 77, 78, 79]. But we note that antibacterial effects have been observed, even when using dried rather than fresh peel, and that pomegranate peel extracts have also been reported to exhibit antiviral [80] and antifungal [75, 76, 79] effects. We were unable to find any relevant data on the *in vivo* effects of these four ingredients with regard to treating soft-tissue wounds or infections.

The most interesting ingredients are undoubtedly honey and breast milk. Here we must add the caveat that honey's sweetness must explain a large part of its ubiquity in pre-modern medicine – a mouthwash containing gall, vinegar, and sumac would certainly be unpalatably bitter in the absence of a sweetening agent! However, the antimicrobial and wound-healing properties of honey have been relatively extensively studied, and honey preparations and dressings are used in clinical settings (e.g. the UK NHS, [81]). Honey's antimicrobial, immunostimulatory and healing effects are attributed to its osmotic properties, an enzyme that produces hydrogen peroxide, bee defensin, phenolics and flavonoids: while the most potent medical-grade honey (manuka honey) also contains highly active compounds from the plant from which it is made, a range of European honeys have been shown to have measurable antimicrobial effects *in vitro* [82, 83, 84, 85]. Four Cochrane reviews find moderate to good evidence that honey reduces healing times for burns and surgical wounds [58, 86], weak evidence for honey helping to prevent mouth ulcers in cancer patients [87] and suggest that honey may be better than no treatment or placebo in relieving childhood cough [88].

Breast milk is similarly recognised as antimicrobial. It contains lysozyme, an enzyme which destroys the structural peptidoglycan layer surrounding bacterial cells, and which is recognised as a key component of innate immunity [89]. It also contains the protein lactoferrin, which sequesters vital iron away from pathogenic bacteria. In acid conditions, lactoferrin is cleaved to procude a peptide called lactoferricin, which has a direct bactericidal effect by binding cell walls and triggering membrane damage [89, 90, 91]. Lactoferrin can kill a range of microbes, including antibiotic-resistant *K. pneumoniae* and *Candida* spp.; it can also enhance the sensitivity of these microbes to modern antibiotics [90, 91]. Clinical research into the antimicrobial potential of breast milk or its component proteins is, however, lacking [92]. It is interesting that donkey milk is suggested as a substitute (*Lylye of Medicynes*, fol. 16v) as the proteome of donkey milk is very similar to that of human milk, and it contains appreciable amounts of lactoferrin and lysozyme [93, 94]. Raw breast milk also contains a range of commensal bacteria, and if milk was not prepared in such a way that these were killed, it is possible that the milk microbiome could contribute to its anti-infection properties. One study found that colostrum could help alleviate infant conjunctivitis because commensal bacteria in the milk outcompeted pathogenic bacteria when applied to the eye [95].

There are several reasons why mixing these ingredients could represent a rational medical decision. First, combining ingredients in a cocktail could increase efficacy against a particular target microbial species by attacking several cellular targets at the same time, or allowing for chemical activation of particular component molecules. For instance, acid pH promotes proteolysis of lactoferrin into peptides called lactoferricins, which have greater antimicrobial activity than the original protein [96]. When milk is directly consumed this happens in the gut and inside neutrophils, but we note that the sumac in *fistula in lacrimali* would decrease pH of the recipe and potentially process lactoferrin into lactoferrcins in the ointment itself. A further possibility is that using multiple ingredients allows for contingency. We note that the available evidence suggests for instance, that Gram-positive bacteria seem to show more sensitivity to sumac than Gram-negative bacteria, whereas the opposite is true for vinegar [67, 69]. Another type of contingency which could be covered by combining multiple antimicrobial ingredients would be protection against potential variation in the composition and quality of particular ingredients, e.g. which *Boswellia* species was the available frankincense sourced from, and did this vary [64]? Finally, we must also consider the physical properties of each ingredient. Mastic, for instance, is not well evidenced as an antimicrobial, but could act as a thickening agent to work other, active, ingredients into an ointment suitable for topical application to an infection site.

## Discussion

Turning a historical medical text into a database suitable for network analysis was achieved through detailed interdisciplinary consideration of the text and its structure. The outcome of this study was dependent on the collaborative efforts and exchange of knowledge between multiple disciplines: medieval studies to translate and contextualise recipes, ingredients, and symptoms; maths and computer science to organise and analyse the resulting database; and microbiology/immunology to determine if "blind" statistically-driven results make sense from an applied perspective.

As described, 'normalizing' or 'cleaning' a medieval data set for analysis with modern computational tools is not a standardized methodology. The process of normalization presented here is likened to a series of editorial decisions, such as preference for a Middle English spelling over a Latin spelling or inclusion of plant parts in the terminology for a base ingredient, but exclusion of preparation instructions. The normalized data set was accomplished by open interdisciplinary dialogue using expertise in medieval language and scientific analytical processing. In documenting our approach to this data set, we hope to add to other ongoing conversations about structuring medieval data in a modern digital space.

This work demonstrates the possibility to use algorithms from complex networks to explore a medieval medical data set for underlying patterns in ingredient combinations related to the treatment of infectious disease. The preliminary results from this pilot study on a single medieval medical text study speak to larger questions regarding medieval medical rationality in constructing treatment recipes, and the application of historical medical knowledge to the current search for novel antimicrobial drugs. Application of this methodology to an expanded data set of ingredients drawn from multiple representative medical and surgical sources may provide a foundation to begin to build evidenced answers to such questions.

**Data Availability Statement**

Our final electronic database of remedies and ingredients in the *Lylye of Medicynes* is provided as **Supplementary Data**. All codes and algorithms used in network analysis are available via C.I.D.G.'s website (https://warwick.ac.uk/fac/cross_fac/complexity/people/staff/delgenio/)

**Acknowledgements**

This work was supported by the Schoenberg Institute for Manuscript Studies, University of Pennsylvania Libraries, the University of Warwick, and the University of Nottingham. We thank the University of Warwick, Institute of Advanced Study for a Residential Fellowship award, June 2017. We thank Christina Lee and the wider Ancientbiotics consortium, including Steve Diggle, for valuable support and discussions; and Spyros Angelopoulos, who participated in early discussions of this work.

**Author contributions**

E.C. conceived the project, created the medieval data set from the *Lylye of Medicynes*, and provided interpretation of Middle English and medieval Latin terminology. C.I.D.G. conducted the network analyses. F.H. conducted the literature review and interpreted the suitability of the results for antimicrobial activity. E.C., C.I.D.G., and F.H. interpreted the data and wrote the manuscript. All authors discussed the results and commented on the manuscript.

**Competing interests**

The authors declare that they have no competing interests.

**Materials and Correspondence**

General correspondence to Erin Connelly. Specific queries on network analysis to Charo I. del Genio

# References

1. Watkins F, Pendry B, Corcoran O, Sanchez-Medina A. Anglo-Saxon pharmacopoeia revisited: a potential treasure in drug discovery. *Drug Discov Today* **16**, 1069-1075 (2011).

2. Watkins F, Pendry B, Sanchez-Medina A, Corcoran O. Antimicrobial assays of three native British plants used in Anglo-Saxon medicine for wound healing formulations in 10th century England. *J Ethnopharmacol* **144**, 408-415 (2012).

3. Harrison F, Roberts AE, Gabrilska R, Rumbaugh KP, Lee C, Diggle SP. A 1,000-Year-Old antimicrobial remedy with antistaphylococcal activity. *MBio* **6**, e01129 (2015).

4. Cameron ML. *Anglo-Saxon Medicine*. Cambridge University Press (2008).

5. Horden P. What's wrong with early medieval medicine? *Social History of Medicine* **24**, 2-25 (2000).

6. Botta F, del Genio CI. Analysis of the communities of an urban mobile phone network. *PLOS One* **12**,  (2017).

7. Farine DR, C.J. Garroway, and B.C. Sheldon. Social network analysis of mixed-species flocks: Exploring the structure and evolution of interspecific social behaviour. *Animal Behaviour* **84**, 1271-1277 (2012).

8. Newman MEJ. The structure of scientific collaboration networks. *Proceedings of the National Academy of Sciences* **98**, 404-409 (2001).

9. Wolf JBW, D. Mawdsley, F. Trillmich, and R. James. Social structure in a colonial mammal: unravelling hidden structural layers and their foundations by network analysis. *Animal Behaviour* **74**, 1293-1302 (2007).

10. Albert R, Jeong H, Barabási A-L. Diameter of the World-Wide Web. *Nature* **401**, 130 (1999).

11. Treviño III S, Sun Y, Cooper TF, Bassler KE. Robust detection of hierarchical communities from *Escherichia coli* gene expression data. *PLOS Computational Biology* **8**,  (2012).

12. Ahn YY, Ahnert SE, Bagrow JP, Barabasi AL. Flavor network and the principles of food pairing. *Sci Rep* **1**, 196 (2011).

13. Albert R, & A.-L. Barabási. Statistical mechanics of complex networks. *Reviews of Modern Physics* **74**, (2002).

14. Boccaletti S, V. Latora, Y. Moreno, M. Chavez and D.-U. Hwang. Complex networks: Structure and dynamics. *Physics Reports* **424**, 175-308 (2006).

15. Boccaletti S, G. Bianconi, R. Criado, C. I. del Genio, J. Gómez-Gardeñes,, M. Romance IS-N, Z. Wang and M. Zanin. The structure and dynamics of multilayer networks. *Physics Reports* **544**, (2014).

16. Newman MEJ. The structure and function of complex networks. *SIAM* **45**, (2003).

17. Barabasi AL, Albert R. Emergence of scaling in random networks. *Science* **286**, 509-512 (1999).

18. del Genio CI, T. Gross and K. E. Bassler. All scale-free networks are sparse. *Physics Review Letters* **107**, (2011).

19. Johnson S, J. J. Torres, J. Marro and M. A. Muñoz. The entropic origin of disassortativity in complex networks. *Physics Review Letters* **104**, (2010).

20. Williams O, del Genio CI. Degree correlations in directed scale-free networks. *PLOS One* **9**, (2014).

21. del Genio CI, M. Romance, R. Criado and S. Boccaletti. Synchronization in dynamical networks with unconstrained structure switching. *Physical Review E* **92**, (2015).

22. del Genio CI, J. Gómez-Gardeñes, I. Bonamassa and S. Boccaletti. Synchronization in networks with multiple interaction layers. *Science Advances* **2**, (2016).

23. Botta F, del Genio CI. Finding network communities using modularity density. *Journal of Statistical Mechanics: Theory and Experiment*, (2016).

24. Fortunato S. Community detection in graphs. *Physics Reports* **486**, 75-174 (2010).

25. Newman MEJ. Communities, modules and large-scale structure in networks. *Nature Physics* **8**, 25-31 (2012).

26. Zachary WW. An Information Flow Model for Conflict and Fission in Small Groups. *Journal of Anthropological Research* **33**, 452-473 (1977).

27. Gleiser PM, Danon L. Community structure in jazz. *Advances in Complex Systems* **06**, 565-573 (2003).

28. Guimerà R, Nunes Amaral LA. Functional cartography of complex metabolic networks. *Nature* **433**, 895 (2005).

29. Huss M, Holme P. Currency and commodity metabolites: their identification and relation to the modularity of metabolic networks. *IET Systems Biology* **1**, 280-285 (2007).

30. Chen M, Kuzmin K, Szymanski BK. Community detection via maximization of modularity and its variants. *IEEE Transactions on Computational Social Systems* **1**, 46-65 (2014).

31. M. C, T. N, K. SB. A new metric for quality of network community structure. *HUMAN* **2**, 226–240 (2013).

32. Treviño III S, Nyberg A, Del Genio C, I., Bassler KE. Fast and accurate determination of modularity and its effect size. *Journal of Statistical Mechanics: Theory and Experiment* **2015**, P02003 (2015).

33. Kovac I*, et al.* Plantago lanceolata L. water extract induces transition of fibroblasts into myofibroblasts and increases tensile strength of healing skin wounds. *J Pharm Pharmacol* **67**, 117-125 (2015).

34. Thomé RG*, et al.* Evaluation of healing wound and genotoxicity potentials from extracts hydroalcoholic of *Plantago major* and *Siparuna guianensis*. *Exp Biol Med (Maywood)* **237**, 1379-1386 (2012).

35. Connelly E. A Case Study of Plantago in the Treatment of Infected Wounds in the Middle English Translation of Bernard of Gordon's *Lilium medicinae*. In: *New Approaches to Disease, Disability, and Medicine in Medieval Europe, Studies in Early Medicine series,* (ed^(eds Connelly E, Künzel S). Archaeopress (in press).

36. Brennessel B, M.D.C. Drout, and R. Gravel. A reassessment of the efficacy of Anglo-Saxon medicine. *Anglo-Saxon England* **34**, 183-195 (2005).

37. Demaitre L. *Doctor Bernard de Gordon: Professor and Practitioner* Pontifical Institute of Mediaeval Studies (1980).

38. Demaitre L. *Medieval Medicine: The Art of Healing from Head to Toe* Praeger (2013).

39. Connelly E. My written books of surgery in the Englishe tonge: the barber-surgeons guild of London and the Lylye of Medicynes. *Manuscript Studies, A Journal of the Schoenberg Institute for Manuscript Studies* **3**,  (2017).

40. Demaitre L. Translations of Bernard of Gordon's Lilium Medicinae, 'A booke practike to meke men'?  (2011).

41. Voigts L. The Master of the King's Stillatories. In: *The Lancastrian Court: Proceedings of the 2001 Harlaxton Symposium* (ed^(eds Stratford J). Shaun Tyas (2003).

42. Baron A, Rayson P. VARD 2: A tool for dealing with spelling variation in historical corpora. *Proceedings of the Postgraduate Conference in Corpus Linguistics, Aston University, Birmingham, UK, 22 May 2008 http://ucrellancsacuk/vard/*, (2008).

43. Lehto A, Baron A, Ratia M, Rayson P. Improving the precision of corpus methods: The standardized version of Early Modern English Medical Texts. In: *Early Modern English Medical Texts: Corpus description and studies* (ed^(eds Taavitsainen I, Pahta P). John Benjamins (2010).

44. Archer D, Kytö M, Baron A, Rayson P. Guidelines for normalising Early Modern English corpora: Decisions and justifications. *ICAME Journal* **39**, 5-24 (2015).

45. *Middle English Dictionary Online*, University of Michigan http://quod.lib.umich.edu/m/med/.

46. Long JR. Botany. In: *Medieval Latin: An Introduction and Bibliographical Guide* (eds F. A. C. Mantello and A. G. Rigg). The Catholic University of America Press (1996).

47. *Oxford English Dictionary, http://www.oed.com*. Oxford University Press.

48. Norri J. *(2016) Dictionary of Medical Vocabulary in English, 1375–1550 Body Parts, Sicknesses, Instruments, and Medicinal Preparations*. Routledge (2016).

49. Medicinal Plant Names Services Portal, Royal Botanic Gardens, Kew. http://mpns.kew.org/mpns-portal/.

50. The Plant List (2010). Version 1. http://www.theplantlist.org/.

51. Simas T, M. Ficek, A. Díaz-Guilera, P. Obrador and P. R. Rodriguez. Food-bridging: a new network construction to unveil the principles of cooking. *Frontiers in ICT* **4**, (2017).

52. Simas TaLMR. Distance closures on complex networks. *Network Science* **3**, 227-268 (2015).

53. Cuthill E, McKee J. Reducing the bandwidth of sparse symmetric matrices. *ACM '69 Proceedings of the 1969 24th National Conference*, 157-172 (1969).

54. Costerton JW, Stewart PS, Greenberg EP. Bacterial biofilms: a common cause of persistent infections. *Science* **284**, 1318-1322 (1999).

55. Reynolds T, Dweck AC. Aloe vera leaf gel: a review update. *J Ethnopharmacol* **68**, 3-37 (1999).

56. Sánchez-Machado DI, J. López-Cervantes, R. Sendón, and A. Sanches-Silva. *Aloe vera*: Ancient knowledge with new frontiers. *Trends in Food Science and Technology* **61**, 94-102 (2017).

57. Dat AD, Poon F, Pham KB, Doust J. *Aloe vera* for treating acute and chronic wounds. *Cochrane Database Syst Rev*, CD008762 (2012).

58. Norman G*, et al.* Antiseptics for burns. *Cochrane Database Syst Rev* **7**, CD011821 (2017).

59. Koutsoudaki C, Krsek M, Rodger A. Chemical composition and antibacterial activity of the essential oil and the gum of *Pistacia lentiscus* Var. chia. *J Agric Food Chem* **53**, 7681-7685 (2005).

60. Aksoy A, Duran N, Koksal F. In vitro and in vivo antimicrobial effects of mastic chewing gum against *Streptococcus mutans* and mutans streptococci. *Arch Oral Biol* **51**, 476-481 (2006).

61. Aksoy A, Duran N, Toroglu S, Koksal F. Short-term effect of mastic gum on salivary concentrations of cariogenic bacteria in orthodontic patients. *Angle Orthod* **77**, 124-128 (2007).

62. Koychev S, Dommisch H, Chen H, Pischon N. Antimicrobial Effects of mastic extract against oral and periodontal pathogens. *J Periodontol* **88**, 511-517 (2017).

63. Sakagami H*, et al.* Selective antibacterial and apoptosis-modulating activities of mastic. *In Vivo* **23**, 215-223 (2009).

64. Moussaieff A, Mechoulam R. *Boswellia* resin: from religious ceremonies to medical uses; a review of in-vitro, in-vivo and clinical trials. *J Pharm Pharmacol* **61**, 1281-1293 (2009).

65. Van Vuuren SF, G.P.P. Kamatou, and A.M. Viljoen. Volatile composition and antimicrobial activity of twenty commercial frankincense essential oil samples. *South African Journal of Botany* **76**, 686-691 (2010).

66. Bjarnsholt T*, et al.* Antibiofilm properties of acetic acid. *Advances in Wound Care* **4**, 363-372 (2015).

67. Halstead FD*, et al.* The Antibacterial Activity of acetic acid against biofilm-producing pathogens of relevance to burns patients. *PLoS One* **10**, e0136190 (2015).

68. A clinical trial looking at the efficacy and optimal dose of acetic acid in burn wound infections ISRCTN11636684 DOI: 10.1186/ISRCTN11636684.

69. de Castro RD, A.C.L.G. Mota, E. de Oliveira Lima, A.U.D. Batista, J. de Araújo Oliveira, and A.L. Cavalcanti. Use of alcohol vinegar in the inhibition of *Candida* spp. and its effect on the physical properties of acrylic resins. *BMC Oral Health* **15**, 52 (2015).

70. Nasar-Abbas SM, Halkman AK. Antimicrobial effect of water extract of sumac (*Rhus coriaria* L.) on the growth of some food borne bacteria including pathogens. *Int J Food Microbiol* **97**, 63-69 (2004).

71. Rayne S, and G. Mazza. Biological activities of extracts from sumac (*Rhus* spp.): A review. *Plant Foods for Human Nutrition* **62**, 165-175 (2007).

72. Hofmann AF, and L. Eckmann. How bile acids confer gut mucosal protection against bacteria. *Proc Natl Acad Sci USA* **103**, 4333-4334 (2006).

73. Sannasiddappa TH, Lund PA, Clarke SR. In vitro antibacterial activity of unconjugated and conjugated bile salts on *Staphylococcus aureus*. *Front Microbiol* **8**, 1581 (2017).

74. Al-Zoreky NS. Antimicrobial activity of pomegranate (*Punica granatum* L.) fruit peels. *Int J Food Microbiol* **134**, 244-248 (2009).

75. Barathikannan K*, et al.* Chemical analysis of *Punica granatum* fruit peel and its *in vitro* and *in vivo* biological properties. *BMC Complement Altern Med* **16**, 264 (2016).

76. Hayouni EA*, et al.* Hydroalcoholic extract based-ointment from *Punica granatum* L. peels with enhanced *in vivo* healing potential on dermal wounds. *Phytomedicine* **18**, 976-984 (2011).

77. Khan I, H. Rahman, N.M. Abd El-Salam, A. Tawab, A. Hussain, T.A. Khan, U.A. Khan, M. Qasim, M. Adnan, A. Azizullah, W. Murad, A. Jalal, N. Muhammad, and R. Ullah. *Punica granatum* peel extracts: HPLC fractionation and LC MS analysis to quest compounds having activity against multidrug resistant bacteria. *BMC Complementary and Alternative Medicine* **17**, 247 (2017).

78. Li G*, et al.* Punicalagin inhibits *Salmonella* virulence factors and has anti-quorum-sensing potential. *Appl Environ Microbiol* **80**, 6204-6211 (2014).

79. de Oliveira JR*, et al.* Cytotoxicity of Brazilian plant extracts against oral microorganisms of interest to dentistry. *BMC Complementary and Alternative Medicine* **13**, 208 (2013).

80. Haidari M, Ali M, Ward Casscells S, 3rd, Madjid M. Pomegranate (*Punica granatum*) purified polyphenol extract inhibits influenza virus and has a synergistic effect with oseltamivir. *Phytomedicine* **16**, 1127-1136 (2009).

81. British National Formulary: https://bnf.nice.org.uk/wound-management/honey-based-topical-application.html; https://bnf.nice.org.uk/wound-management/sheet-dressing.html

82. Fidaleo M, A. Zuorro, and R. Lavecchia. Antimicrobial activity of some italian honeys against pathogenic bacteria. *Chemical Engineering Transactions*, 1015-1020 (2011).

83. Grego E*, et al.* Evaluation of antimicrobial activity of Italian honey for wound healing application in veterinary medicine. *Schweiz Arch Tierheilkd* **158**, 521-527 (2016).

84. Molan PC. Potential of honey in the treatment of wounds and burns. *Am J Clin Dermatol* **2**, 13-19 (2001).

85. Roberts A, H. Brown, and R. Jenkins. On the antibacterial effects of manuka honey: mechanistic insights. *Research and Reports in Biology* **6**, 215-244 (2015).

86. Jull AB, N. Cullum, J.C. Dumville, M.J. Westby, S. Deshpande, and N. Walker. Honey as a topical treatment for wounds. *Cochrane Database of Systematic Reviews* **3**, (2015).

87. Worthington HV*, et al.* Interventions for preventing oral mucositis for patients with cancer receiving treatment. *Cochrane Database Syst Rev*, CD000978 (2011).

88. Oduwole O, Meremikwu MM, Oyo-Ita A, Udoh EE. Honey for acute cough in children. *Cochrane Database Syst Rev*, CD007094 (2014).

89. Cacho NT, Lawrence RM. Innate immunity and breast milk. *Frontiers in Immunology* **8**, 584 (2017).

90. Fernandes KE, Carter DA. The antifungal activity of lactoferrin and its derived peptides: mechanisms of action and synergy with drugs against fungal pathogens. *Frontiers in Microbiology* **8**, 2 (2017).

91. Morici P*, et al.* Synergistic activity of synthetic N-terminal peptide of human lactoferrin in combination with various antibiotics against carbapenem-resistant *Klebsiella pneumoniae* strains. *Eur J Clin Microbiol Infect Dis*, (2017).

92. Pammi M, Abrams SA. Oral lactoferrin for the treatment of sepsis and necrotizing enterocolitis in neonates. *Cochrane Database Syst Rev*, CD007138 (2011).

93. Piovesana S*, et al.* Peptidome characterization and bioactivity analysis of donkey milk. *J Proteomics* **119**, 21-29 (2015).

94. Tidona F, C. Sekse, A. Criscione, M. Jacobsen, S. Bordonaro, D. Marletta, and G.E. Vegarud. Antimicrobial effect of donkeys' milk digested in vitro with human gastrointestinal enzymes. *International Dairy Journal* **21**, 158-165 (2011).

95. Singh M, Sugathan PS, Bhujwala RA. Human colostrum for prophylaxis against sticky eyes and conjunctivitis in the newborn. *J Trop Pediatr* **28**, 35-37 (1982).

96. Gifford JL, Hunter HN, Vogel HJ. Lactoferricin: a lactoferrin-derived peptide with antimicrobial, antiviral, antitumor and immunological properties. *Cell Mol Life Sci* **62**, 2588-2598 (2005).